\documentclass{article}
\usepackage{spconf,amsmath,graphicx}
\usepackage{amssymb,amsfonts}
\usepackage{algorithm}
\usepackage{algpseudocode}
\usepackage{breqn}
\usepackage{mathtools}
\usepackage{tikz}
\usetikzlibrary{shapes,arrows}
\usepackage{subcaption}
\usepackage{cite}
\usepackage{xcolor,soul}
\usepackage{hyperref}
\usepackage{siunitx}
\input{mysymbol.sty}
\newtheorem{problem}{\hspace{0pt}\bf Problem}

\renewcommand{\baselinestretch}{0.97} 

\DeclareSIUnit{\symbol}{\text{symb}}

\title{Accelerated massive MIMO detector based on annealed underdamped Langevin dynamics}
\name{Nicolas Zilberstein$^{\star}$, Chris Dick$^{\dagger}$, Rahman Doost-Mohammady$^{\star}$, Ashutosh Sabharwal$^{\star}$, Santiago Segarra$^{\star}$\thanks{This work was partially supported by Nvidia. Email: \{nzilberstein, doost, ashu, segarra\}@rice.edu, cdick@nvidia.com.}}
\address{$^{\star}$Rice University, USA \hspace{4cm}
         $^{\dagger}$Nvidia, USA}
         
\begin{document}
\ninept
\maketitle
\begin{abstract}


We propose a multiple-input multiple-output (MIMO) detector based on an annealed version of the \emph{underdamped} Langevin (stochastic) dynamic.
Our detector achieves state-of-the-art performance in terms of symbol error rate (SER) while keeping the computational complexity in check.
Indeed, our method can be easily tuned to strike the right balance between computational complexity and performance as required by the application at hand.
This balance is achieved by tuning hyperparameters that control the length of the simulated Langevin dynamic.
Through numerical experiments, we demonstrate that our detector yields lower SER than competing approaches (including learning-based ones) with a lower running time compared to a previously proposed \emph{overdamped} Langevin-based MIMO detector.
\end{abstract}
\begin{keywords}
Massive MIMO detection, Markov chain Monte Carlo, underdamped Langevin dynamics, diffusion process
\end{keywords}
%

\section{Introduction}\label{S:intro}

Massive multiple-input multiple-output (MIMO) systems play a key role in the development of modern and future communications~\cite{mimoreview1}.
In essence, base stations in massive MIMO systems are equipped with a large number of receiver antennas, enabling them to handle several users simultaneously.
This feature enables higher data rates and spectral efficiency, which are fundamental for moving from the fifth to the sixth generation of cellular communications~\cite{6g}.
However, these systems entail many challenges such as designing low-complexity MIMO detection schemes, which is our focus.

Exact MIMO detection is an NP-hard problem~\cite{Pia2017MixedintegerQP}. 
Given $N_u$ users and a modulation of $K$ symbols, the exact maximum likelihood (ML) estimator has an exponential decoding complexity $\mathcal{O}(K^{N_u})$.
Thus, the combination of a large number of users and higher order modulation schemes makes the ML estimate computationally intractable even for moderately-sized systems. 
Many approximate solutions for symbol detection have been proposed in the classical literature including zero-forcing (ZF) and minimum mean squared error (MMSE)~\cite{Proakis2007}. 
Both are (linear) low-complexity detectors but their performance degrades severely for larger systems~\cite{chockalingam_rajan_2014}.
Another classical detector is approximate message passing (AMP), which is asymptotically optimal for large MIMO systems with Gaussian channels but degrades significantly for other (more practical) channel distributions~\cite{amp}. 
In the past few years, several massive MIMO symbol detectors based on machine learning -- and, in particular, deep learning -- have been derived, which can be roughly categorized into channel-specific methods, like MMNet~\cite{mmnet} and channel-agnostic methods like RE-MIMO~\cite{remimo}, OAMPNet~\cite{oampnet}, and hyperMIMO~\cite{hypermimo,zilberstein2021robust}.

An alternative family of detectors is based on Markov chain Monte Carlo (MCMC) methods~\cite[Chapter~8]{chockalingam_rajan_2014}.
Given that the ML estimator is prohibitive for large systems, these methods seek a solution by generating candidate samples from the search space.
Recently, the annealed version of the \emph{overdamped} (or first-order) Langevin dynamic has been used for inverse problems in different areas, such as in image processing (denoising, inpainting)~\cite{kawar2021snips} and communications~\cite{zilberstein2022annealed}, achieving state-of-the-art results.
In essence, the Langevin dynamic explores a target distribution by moving in the direction of the gradient of the logarithm of a target density with an additional noise term, thus allowing the method to avoid collapsing to local maxima.
Although this method shows impressive results, the non-asymptotic convergence can be slow, in particular for real-time applications.
Therefore, in the past few years, the \emph{underdamped} (or \emph{second-order}) Langevin dynamic has gained interest as it has shown a better convergence rate in the non-asymptotic regime compared to the first-order case~\cite{pmlr-v75-cheng18a}.
In a nutshell, a momentum variable is added to the dynamic, which entails smoothing trajectories and thus improves the mixing time, an effect that resembles acceleration in classical gradient descent~\cite{YiAn2021IsTA}.

Given the better non-asymptotic convergence rate of the {underdamped} dynamic, in this paper we propose a general framework to solve linear inverse problems using an annealed version of the \emph{underdamped} Langevin dynamic, and we apply it to the problem of MIMO detection.
Thus, we seek a solution that strikes a balance between state-of-the-art performance and low-running time complexity.
Moreover, we incorporate the annealing process to the {underdamped} Langevin dynamic, which allows us to include the prior information about the discrete nature of the signal of interest (constellation symbols in MIMO detection).

\vspace{0.2mm}
\noindent
{\bf Contribution.}
The contributions of this paper are twofold:\\
1) We propose a general framework for solving linear inverse problems based on \emph{annealed underdamped Langevin dynamics}, allowing us to reduce the computational complexity compared to the {overdamped} case, and we apply to the problem of MIMO detection.\\
2) Through numerical experiments, we analyze the behavior of our method for different hyperparameter settings and demonstrate that the proposed detector achieves state-of-the-art symbol error rate (SER) with low running time.

\vspace{-2mm}
\section{System model and problem formulation}


The forward model for the MIMO system given $N_u$ single-antenna transmitters or users and a receiving base station with $N_r$ antennas is defined as
\begin{equation}\label{E:mimo_model}
	\bby = \bbH \bbx + \bbz,
\end{equation}
where $\bbH \in \mathbb{C}^{N_r \times N_u}$ is the channel matrix, $\bbz \sim \mathcal{CN}(\bb0, \sigma_0^2 \bbI_{N_r})$ is a vector of complex circular Gaussian noise, $\bbx \in \mathcal{X}^{N_u}$ is the vector of transmitted symbols, $\mathcal{X}$ is a finite set of constellation points, and $\bby \in \mathbb{C}^{N_r}$ is the received vector. 
We consider a quadrature amplitude modulation (QAM) throughout this work and symbols are normalized to attain unit average power. 
All the users transmit with the same modulation and each symbol has the same probability of being chosen by the users $N_{u}$. 
Moreover, we assume perfect channel state information (CSI), which means that $\bbH$ and $\sigma_0^2$ are known at the receiver.\footnote{To avoid notation overload, we adopt the convention that whenever we assume $\bbH$ to be known, $\sigma_0^2$ is also known.} 
Under this configuration, the MIMO detection problem can be {stated} as follows.
\vspace{-0.03in}
\begin{problem}\label{P:main} \emph{
	Given perfect CSI and an observed $\bby$ following~\eqref{E:mimo_model}, find an estimate of $\bbx$.}
\end{problem}
\vspace{-0.03in}
Given that $\bbz$ in~\eqref{E:mimo_model} is a random variable, a natural way of solving Problem~\ref{P:main} is to search for the $\bbx$ that maximizes its \emph{posterior} probability given the noisy observations $\bby$.
Hence, the Bayes' optimal decision rule can be written as
\begin{align}\label{eq:map}
	\hat{\bbx}_{\mathrm{MAP}} = \argmax_{\bbx \in \mathcal{X}^{N_u}}\,\, p(\bbx|\bby,\bbH)
	= \argmax_{\bbx \in \mathcal{X}^{N_u}}\,\, p_{\bbz}(\bby - \bbH\bbx)p(\bbx).
\end{align}
%
As we assume that the symbols' prior distribution is uniform among the constellation elements and the measurement noise $\bbz$ is Gaussian, the maximum a posteriori (MAP) detector boilds down to an ML detector. 
Specifically,~\eqref{eq:map} is equivalent to the following optimization problem
\begin{equation}\label{eq:ml}
	\hat{\bbx}_{\mathrm{ML}} = \argmin_{\bbx \in \mathcal{X}^{N_u}}\,\, ||\bby - \bbH\bbx||^2_2,
\end{equation}
%
\noindent which is NP-hard due to the finite constellation constraint $\bbx \in \mathcal{X}^{N_u}$, rendering $\hat{\bbx}_{\mathrm{ML}}$ intractable in practical applications.
Thus, several schemes have been proposed in the last decades to provide efficient approximate solutions to Problem~\ref{P:main}, as mentioned in Section~\ref{S:intro}.
In this paper, we propose to solve Problem~\ref{P:main} by (approximately) sampling from the posterior distribution in~\eqref{eq:map} using an annealed {underdamped} Langevin dynamic.

\vspace{-0.1in}
\section{Underdamped Langevin for MIMO detection}
\vspace{-0.07in}

In Section~\ref{subsec:langevindyn}, we briefly introduce the underdamped Langevin dynamic, also known as second-order Langevin dynamic, while in Section~\ref{subsec:discretization} we explain the numerical implementation of the continuous-time dynamic based on the splitting method.
Finally, in section~\ref{subsec:posterior} we detail the expressions of the score functions involved in the sampling process to solve our Problem~\ref{P:main}.
\vspace{-0.1in}
\subsection{Underdamped Langevin dynamics}
\label{subsec:langevindyn}

The {underdamped} Langevin diffusion is the Markov process on variables $\bbx_t \in \mathbb{R}^d$ and $\bbv_t \in \mathbb{R}^d$ that solves the stochastic differential equations (SDEs)
\begin{align}\label{eq:ct_underdamped_langevin}
    \text{d}\bbv_t &= -\nabla U(\bbx_t) - \gamma \bbv_t\text{d}t + \sqrt{2 \gamma \tau} \bbM^{-\frac{1}{2}}\text{d}\bbW,\\
    \text{d}\bbx_t &= \bbM^{-1}\bbv_t \text{d}t,\nonumber
\end{align}
\noindent where $\bbW$ is a standard $d$-dimensional Brownian motion, $U \in \ccalC^2(\mathbb{R}^d)$ is called the potential, $\gamma>0$ is a friction parameter, $\bbM$ is a mass matrix that controls the coupling between $\bbx_t$ and $\bbv_t$, and $\tau$ is a temperature parameter.
Under mild conditions, it can be shown that the invariant distribution of the continuous-time process is $\pi(\bbx, \bbv) \propto \exp[- (\tau^{-1}U(\bbx) + \frac{\bbv^\top \bbM^{-1}\bbv}{2})]$~\cite{pavliotis_book}.
Given a target distribution $p(\bbx)$ from which we want to generate samples $\bbx \in \mathbb{R}^d$, if we define $U(\bbx) = -\log p(\bbx)$, then~\eqref{eq:ct_underdamped_langevin} defines an MCMC sampler as  $\pi(\bbx) \propto p(\bbx)^{1/\tau}$.
In particular, if $ \tau = 1$, then $\pi(\bbx) \propto p(\bbx)$.

The \emph{overdamped} Langevin dynamic can be obtained as a particular regime of the dynamic in~\eqref{eq:ct_underdamped_langevin}, when the friction parameter $\gamma \xrightarrow{} \infty$~\cite[Section~6.5]{pavliotis_book}.
In essence, the dynamic in~\eqref{eq:ct_underdamped_langevin} explores the target distribution by moving in the direction of the gradient of the logarithm of the target density $\nabla_{\bbx}\log p(\bbx)$, known as \emph{score function}.
Through the lens of sampling, the momentum term entails, under some assumptions on the log-probability density $\log p(\bbx)$, an accelerated version of the sampler compared to the overdamped Langevin dynamic.
In particular, the~\cite{pmlr-v75-cheng18a} establishes a non-asymptotic convergence rate for a specific discretization scheme of~\eqref{eq:ct_underdamped_langevin} when $\log p(\bbx)$ is {strongly concave} and has Lipschitz continuous gradient (this result was later improved in~\cite{Dalalyan2018OnSF}).
In~\cite{Monmarche2021HighdimensionalMW}, another convergence rate is given by considering a splitting strategy as the discretization scheme; details about discretization schemes are postponed to Section~\ref{subsec:discretization}.
Overall, these results demonstrate that there is a significant improvement when considering the underdamped case.

\vspace{-0.1in}
\subsection{Splitting method as a discretization scheme}
\label{subsec:discretization}

In this work, we rely on splitting methods for the discretization of~\eqref{eq:ct_underdamped_langevin}. 
In a nutshell, given the operator that describes the time evolution of the state $\bbx_t$, the idea is to split the operator into tractable sub-operators and then compose them to approximate the full operator. 
Formally, given an initial state $\bbx_0$, the solution of the SDE~\eqref{eq:ct_underdamped_langevin} can be constructed as $\bbx_t = \exp{(t\ccalL)} \bbx_0$, where $\exp(t\ccalL)$ is the operator that defines the propagation of the state and $\ccalL$ is the (infinitesimal) generator of the Markov process~\cite{pavliotis_book}. 
Then, we split the operator $\ccalL$ into sub-operators as explained in~\cite{Leimkuhler}, and rewrite~\eqref{eq:ct_underdamped_langevin} as 
%

{\footnotesize{
\begin{align}
    \hspace{-1mm}\text{d}\begin{bmatrix}\bbv_t \\ \bbx_t \end{bmatrix}= \underbrace{\begin{bmatrix}\bb0 \\ \bbM^{-1}\bbv_t \end{bmatrix}\text{d}t}_{\text{H}_1} + \underbrace{\begin{bmatrix}\nabla\log p(\bbx) \\ \bb0 \end{bmatrix}\text{d}t}_{\text{H}_2} + \underbrace{\begin{bmatrix} - \gamma \bbv_t\text{d}t + \sqrt{2 \gamma \tau} \bbM^{-\frac{1}{2}}\text{d}\bbW \\ \bb0 \end{bmatrix}}_{\text{O}}. \nonumber
\end{align}}}

\noindent \normalsize The label $\text{H}_1$ and $\text{H}_2$ refer to the Hamiltonian components, which can be solved using any deterministic numerical integrator, and the label O refers to the Ornstein-Uhlenbeck process, {which has a closed-form expression when integrating it in an interval $[kh, (k+1)h)$}.
Therefore, given a step size $\epsilon$, the discretization scheme is given by
%
{\small\begin{align}\label{eq:discrete_langevin}
\bbx_{k+1} &= \bbx_{k} + \epsilon \bbM^{-1}\bbv_{k}, \\
\nonumber \bbv_{k+1/2} &= \bbv_{k} + \epsilon\nabla\log p(\bbx_{k+1}), \\
\nonumber \bbv_{k+1} &= \exp(-\gamma \epsilon)\bbv_{k+1/2} + \sqrt{\tau(1-\exp(-2\gamma \epsilon))}\bbM^{\frac{1}{2}}\bbw_k,
\end{align}}

\noindent \normalsize where $\bbw_k \sim \ccalN(0, \bbI)$.
This particular scheme corresponds to the adjoint symplectic Euler scheme to solve the Newtonian part of the Langevin dynamics SDE, followed by an exact Ornstein-Uhlenbeck solution.
Other integrators for the Hamiltonian part -- like a velocity Verlet numerical integrator -- can be used, giving place to other splitting strategies~\cite{Leimkuhler}.

The numerical integration scheme used to discretize~\eqref{eq:ct_underdamped_langevin} largely determines the performance of the algorithm. 
The simplest scheme is the \emph{Euler-Maruyama discretization}, which is a first-order integrator.
Although its implementation is very simple, this discretization scheme does not grant an accelerated convergence~\cite{YiAn2021IsTA}.
Moreover, in~\cite{Leimkuhler} it is shown that the order of accuracy\footnote{The order of accuracy is the exponent in the power law by which the error in the method is related to the step size.} is lower compared to more sophisticated schemes. 
An alternative scheme is the one proposed in~\cite{pmlr-v75-cheng18a}, which is as follows. 
First, the time dimension in~\eqref{eq:ct_underdamped_langevin} is discretized into intervals of equal length $h$. 
Then, in the $(k + 1)$-th step, a continuous dynamic in the interval $\tau \in [kh, (k + 1)h]$ is defined by conditioning on the initial values $(\bbx_{kh}, \bbv_{kh})$.
Finally, the continuous SDE in this interval is solved, which yields a Gaussian distribution as the solution.
{However, in our implementation we use the splitting method leading to the discretization scheme in~\eqref{eq:discrete_langevin} as it showed better empirical performance.}

\vspace{-0.1in}
\subsection{Sampling from the posterior distribution with acceleration}
\label{subsec:posterior}

{Recall that our goal is to solve Problem~\ref{P:main} by sampling (approximately) from the posterior defined in~\eqref{eq:map} using the underdamped Langevin dynamic~\eqref{eq:discrete_langevin}.
However, notice that the framework developed in Sections~\ref{subsec:langevindyn} and~\ref{subsec:discretization} does not apply directly to Problem~\ref{P:main} for two reasons.
First, \emph{we do not seek to sample from $p(\bbx)$, but rather from the posterior $p(\bbx|\bby, \bbH)$.}
Thus, the score is $\nabla_{\bbx}\log p(\bbx| \bby, \bbH)$, which can be written after applying Bayes' rule as 
\begin{equation}\label{E:score_function}
\nabla_{\bbx}\log p(\bbx|\bby, \bbH) = \nabla_{\bbx}\log p(\bby|\bbx,\bbH) + \nabla_{\bbx}\log p(\bbx),
\end{equation}
where the term $\nabla_{\bbx}\log p(\bby|\bbx,\bbH)$ corresponds to the score function of the likelihood and $\nabla_{\bbx}\log p(\bbx)$ to the score function of the prior.
Second, \emph{the gradient with respect to the prior of $\bbx$ is not well defined as $\bbx$ belongs to a discrete constellation set.}
To circumvent this issue, we propose to leverage an annealing process. }
Therefore, in this section we present our algorithm based on the annealed version of the underdamped Langevin dynamics and give the closed-form expressions of the two terms involved in~\eqref{E:score_function}.
First, we define a sequence of noise levels $\{\sigma_l\}_{l=1}^{L+1}$ such that $\sigma_1 > \sigma_2 > \cdots > \sigma_L > \sigma_{L+1} = 0$. 
Then, at each level we define a perturbed version of the true symbols $\bbx$
\begin{equation}\label{eq:pert_symbs}
    \tilde{\bbx}_{l} = \bbx + \bbn_{l},
\end{equation}
where $\bbn_{l} \sim \mathcal{CN}(0, \sigma_l^2\bbI)$. 
Furthermore, to get tractable expressions, we rely on the singular value decomposition (SVD) of the channel matrix given by $\bbH = \bbU\boldsymbol{\Sigma}\bbV^\top$ as well as in the spectral representation of $\tilde{\bbx}_l$ and $\bby$ defined as $\tilde{\boldsymbol{\chi}}_l = \bbV^{\top} \tilde{\bbx}_l$ and $\boldsymbol{\eta} = \bbU^{\top}\bby$. 
In the spectral domain, the score function for every noise level $l$ is given by [cf.~\eqref{E:score_function}]
\begin{equation}\label{E:score_function_spectral}
\nabla_{\tilde{\boldsymbol{\chi}}_l}\!\log p(\tilde{\boldsymbol{\chi}}_l| \boldsymbol{\eta}, \bbH) = \nabla_{\tilde{\boldsymbol{\chi}}_l} \log p(\boldsymbol{\eta}|\tilde{\boldsymbol{\chi}}_l, \bbH) + \nabla_{\tilde{\boldsymbol{\chi}}_l}\log p(\tilde{\boldsymbol{\chi}}_l).
\end{equation}
We now provide closed-form expressions for both constituent terms in this score function.
\vspace{1mm}

\noindent \emph{i) Score of the likelihood:} The final expression for the score of the likelihood in the spectral domain is given by
\begin{equation}\label{eq:score_likeli}
    \nabla_{\tilde{\boldsymbol{\chi}}_l} \! \log p(\boldsymbol{\eta}|\tilde{\boldsymbol{\chi}}_l, \bbH)  = \boldsymbol{\Sigma}^\top \,\, |\sigma_0^2\bbI - \sigma_l^2\boldsymbol{\Sigma}\boldsymbol{\Sigma}^\top|^{\dagger}\,\, ( \boldsymbol{\eta} - \boldsymbol{\Sigma} \tilde{\boldsymbol{\chi}}_l).
\end{equation}
Details of why the score function of the likelihood is given by the gradient of a multivariate Gaussian distribution can be found in~\cite{zilberstein2022annealed}.

\vspace{1mm}
\noindent \emph{ii) Score of the annealed prior:} The score function can be related to the MMSE denoiser through Tweedie's identity~\cite{TweedieIdent} as follows
\begin{equation}\label{eq:prior}
	\nabla_{\tilde{\bbx}_l}\log p(\tilde{\bbx}_l) = \frac{\mathbb{E}_{\sigma_l}[\bbx|\tilde{\bbx}_l] - \tilde{\bbx}_l}{\sigma_l^2}.
\end{equation}
In particular, the conditional expectation can be calculated elementwise as
\begin{align}\label{E:mixed_gaussian}
	\mathbb{E}_{\sigma_l}[x_j|[\tilde{\bbx}_l]_j] &= \frac{1}{Z}\sum_{x_k \in \ccalX} x_k \exp\bigg(\frac{-||[\tilde{\bbx}_l]_j - x_k||^2}{2\sigma_l^2}\bigg),
\end{align}
where $Z = \sum_{x_k \in \ccalX} \exp\Big(\frac{-||[\tilde{\bbx}_l]_j - x_k||^2}{2\sigma_l^2}\Big)$ and $j=1,\cdots, N_u$. 
Then, given the orthogonality of $\bbV$, we have $\nabla_{\tilde{\boldsymbol{\chi}}_l}\log p(\tilde{\boldsymbol{\chi}}_l) = \bbV^{\top}\nabla_{\tilde{\bbx}_l}\log p(\tilde{\bbx}_l)$.

\begin{algorithm}[t]
	\caption{Annealed underdamped Langevin for MIMO detection}\label{alg}
	\begin{algorithmic}
		\Require $T, \{\sigma_l\}_{l=1}^L, \epsilon, \sigma_0, \bbH, \bby, \tau, \bbM$
		\State Compute SVD of $\bbH = \bbU\boldsymbol{\Sigma}\bbV^{\top}$
		\State Initialize $\tilde{\boldsymbol{\chi}}_{t=0,l=1}$ with random noise $\ccalU[-1,1]$
		\For{$l = 1\; \text{to}\;  L$}
		\State $[\boldsymbol{\Lambda}_l]_{jj} =$
		$\begin{cases}
		  \frac{\epsilon \sigma_l^2 }{\sigma_L^2} (1 - \frac{\sigma_l^2}{\sigma_0^2}s_j^2) \hspace{8mm} \text{if} \,\,\, \sigma_ls_j \leq \sigma_0 \\
		\frac{\epsilon}{\sigma_L^2} (\sigma_l^2 - \frac{\sigma_0^2}{s_j^2}) \hspace{10mm}  \text{if} \,\,\, \sigma_ls_j > \sigma_0
		\end{cases}$
		
		\For{$k = 0\; \text{to}\; T-1$}
			\State Draw $\bbw_k \sim \ccalN(0, \bbI)$
			
            \State Compute $\nabla_{\tilde{\boldsymbol{\chi}}_{k,l}}\log p(\boldsymbol{\eta}|\tilde{\boldsymbol{\chi}}_{k,l}, \bbH)$ as in~\eqref{eq:score_likeli}
            
            \State Compute $\nabla_{\tilde{\bbx}_{k,l}}\log p(\tilde{\bbx}_{k,l})$
            as in~\eqref{eq:prior}
            
            \State Compute $\nabla_{\tilde{\boldsymbol{\chi}}_{k,l}}\!\log p(\tilde{\boldsymbol{\chi}}_{k,l}| \boldsymbol{\eta}, \bbH)$ as in~\eqref{eq:full_score}


            \State $\tilde{\boldsymbol{\chi}}_{k+1, l} = \tilde{\boldsymbol{\chi}}_{k,l} + \frac{\epsilon}{\sigma_L^2}\bbM^{-1}\bbv_{t}$
            \State $\bbv_{k+1/2} = \bbv_k + \boldsymbol{\Lambda}_l \nabla_{\tilde{\boldsymbol{\chi}}_{k,l}}\!\log p(\tilde{\boldsymbol{\chi}}_{k,l}| \boldsymbol{\eta}, \bbH)$
            \State $\bbv_{k+1} \! = \! e^{-\gamma {\epsilon}/{\sigma_L^2}}\bbv_{k+1/2} + \sqrt{\tau(1-e^{-2\gamma {\epsilon}/{\sigma_L^2}})}\bbM^{\frac{1}{2}} \boldsymbol{\Lambda}_l\bbw_k$
		\EndFor
		\State $\tilde{\boldsymbol{\chi}}_{0, l+1} = \tilde{\boldsymbol{\chi}}_{T, l}$
		\EndFor \\
	\Return $\bar{\bbx} = \argmin_{\bbx \in \ccalX^{N_u}}||\bbx - \bbV\tilde{\boldsymbol{\chi}}_{T,L}||_2^2$
	\end{algorithmic}
\end{algorithm}

\begin{figure*}[t]
	\begin{subfigure}{.33\textwidth}
    	\centering
    	\includegraphics[width=1\textwidth]{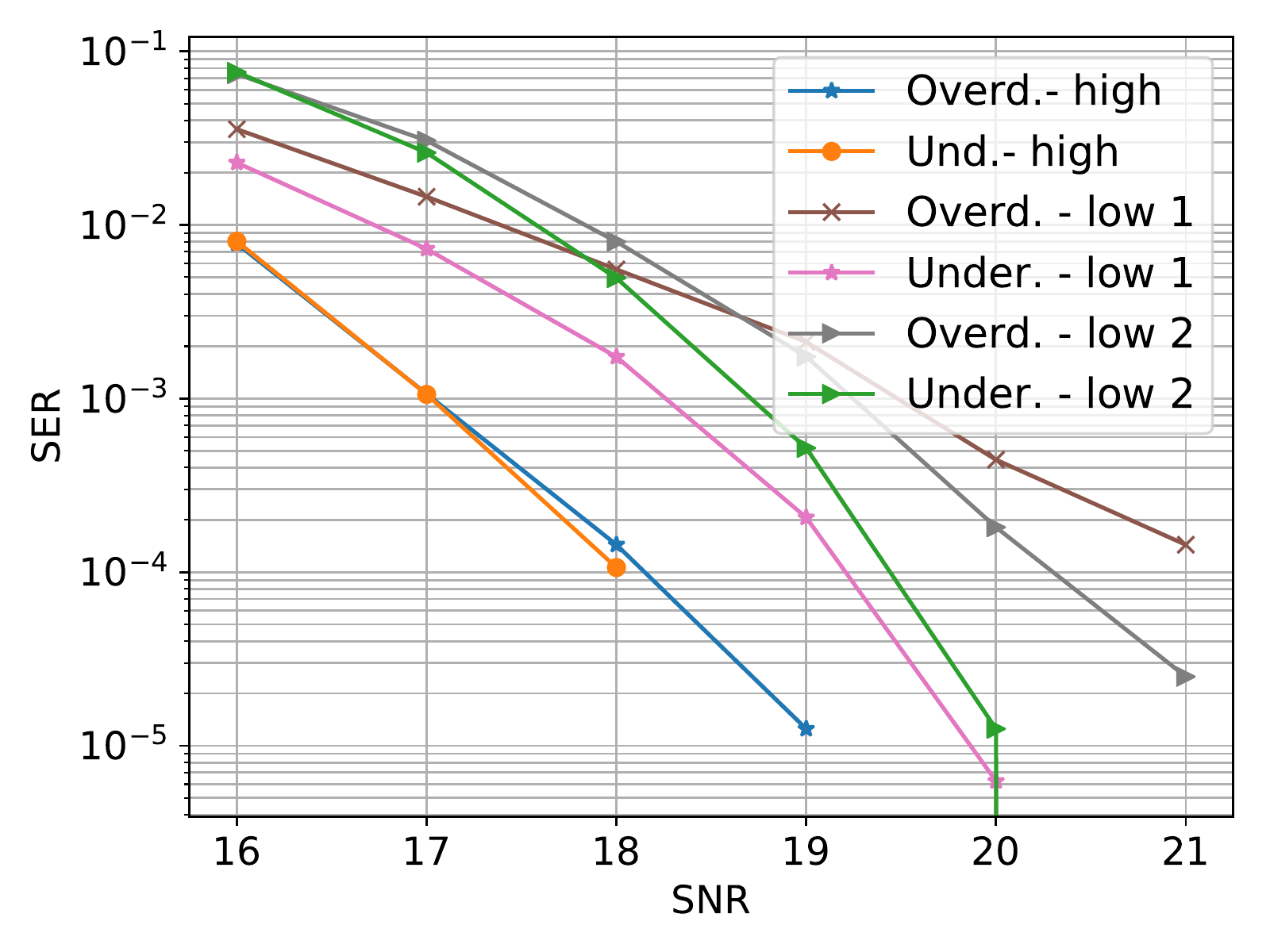}
    	\vspace{-0.2in}
    	\caption{}
    	\label{fig:SER-comparison_order}
	\end{subfigure}
	\begin{subfigure}{.33\textwidth}
    	\centering
    	\includegraphics[width=1\textwidth]{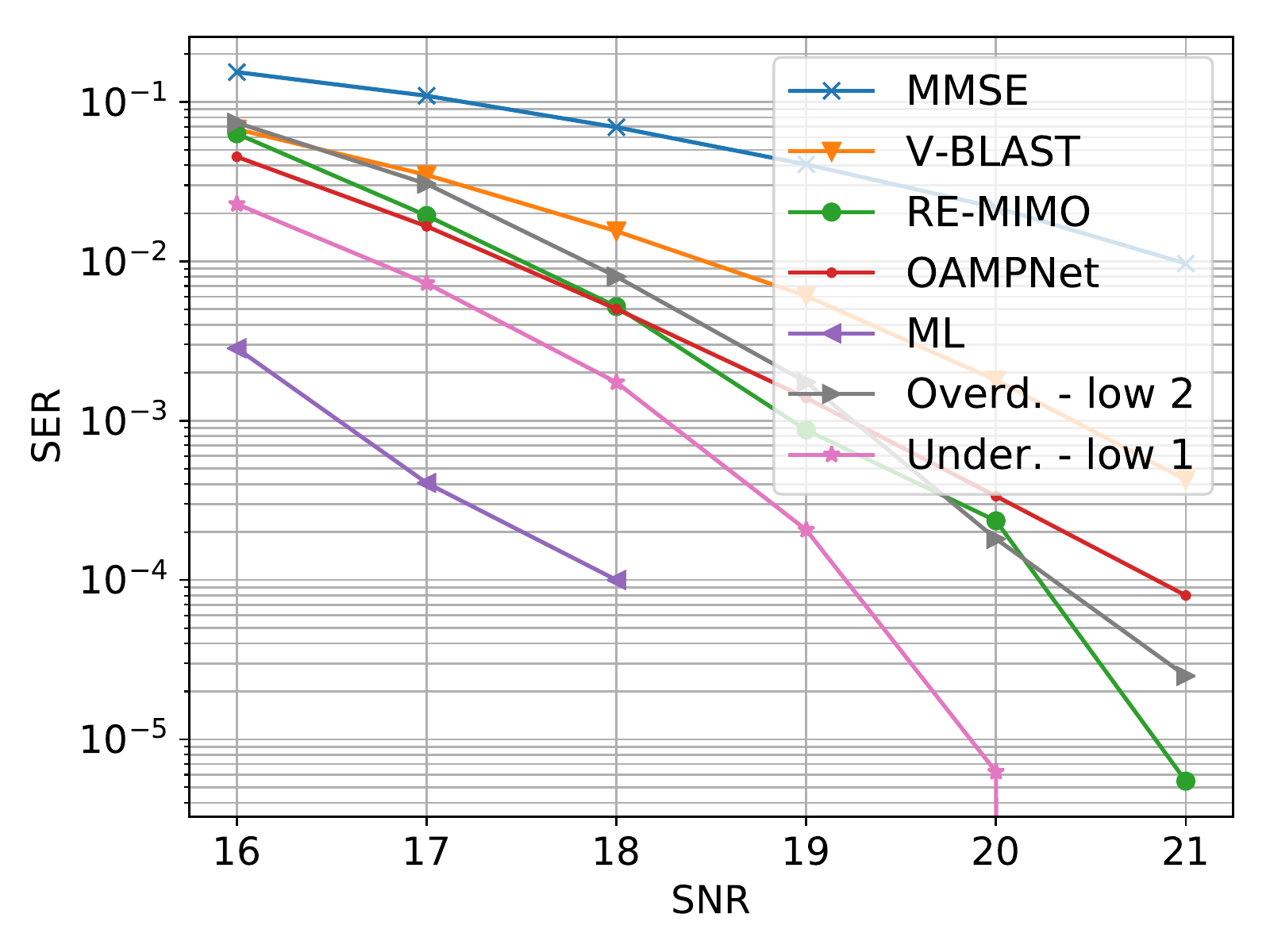}
    	\vspace{-0.2in}
    	\caption{}
    	\label{fig:SER-comparison}
	\end{subfigure}
	\begin{subfigure}{.33\textwidth}
    	\centering
    	\includegraphics[width=1\textwidth]{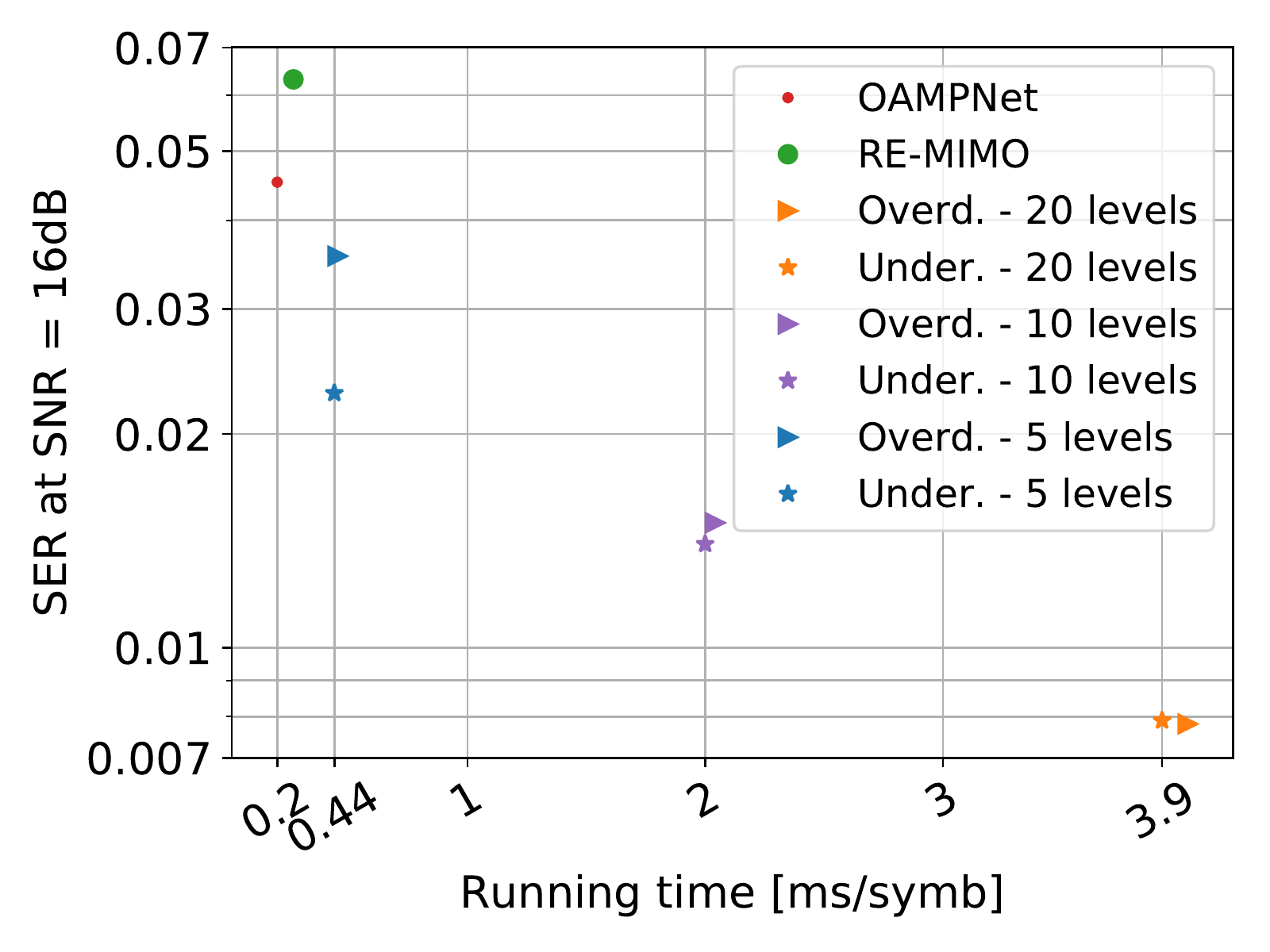}
    	\vspace{-0.2in}
    	\caption{}
    	\label{fig:SER-complexity}
	\end{subfigure}%
	\vspace{-0.08in}
	\caption{ {\small Performance analysis of our proposed method considering SER as function of SNR in a Kronecker correlated channel model as in~\eqref{E:kron}. 
	(a)~Comparison between our underdamped Langevin method and the overdamped variant. 
	(b)~Comparison with classical and learning-based detection methods. 
	(c)~Comparison with different detection methods for an $\text{SNR}=\SI{16}{\decibel}$ and varying the running time in $\mathrm{ms}/\mathrm{symb}$.}}
	\vspace{-0.1in}
	\label{fig_results}
\end{figure*}

\vspace{2mm}
\noindent {\bf Algorithm.} 
The algorithm to generate samples $\hat{\bbx}$ from the (approximate) posterior $p(\bbx|\bby,\bbH)$ is shown in Algorithm~\ref{alg}.
We use a user-dependent step size and we calculate the score elementwise as follows~\cite{zilberstein2022annealed}
 \begin{align}\label{eq:full_score}
 [\nabla_{\tilde{\boldsymbol{\chi}}_l} & \! \log p(\tilde{\boldsymbol{\chi}}_{l}| \boldsymbol{\eta}, \bbH)]_j = \\
 & 
    \begin{cases}
  [\nabla_{\tilde{\boldsymbol{\chi}}_l} \! \log p(\boldsymbol{\eta}|\tilde{\boldsymbol{\chi}}_l, \bbH) + \bbV^{\top}\nabla_{\tilde{\bbx}}\log p(\tilde{\bbx})]_j, \,\,\,\, \sigma_0 \geq \sigma_ls_j,
  \\
  [\nabla_{\tilde{\boldsymbol{\chi}}_l} \! \log p(\boldsymbol{\eta}|\tilde{\boldsymbol{\chi}}_l, \bbH)]_j,  \hspace{27mm} \sigma_0 < \sigma_ls_j, \\
  [\bbV^{\top}\nabla_{\tilde{\bbx}} \! \log p(\tilde{\bbx})]_j, \hspace{33mm}  s_j = 0.
\end{cases}
\nonumber
\end{align}
By comparing Algorithm~\ref{alg} with~\eqref{eq:discrete_langevin} it follows that we are implementing the \emph{momentum term directly in the spectral domain}.
Putting it differently, our velocity variable $\bbv$ is capturing the time derivative of the position in the spectral domain directly.
Additionally, given that the sample after the annealing process will be very close to the constellation but not exactly, we take $\bar{\bbx} = \argmin_{\bbx \in \ccalX^{N_u}}||\bbx - \bbV\tilde{\boldsymbol{\chi}}_{T,L}||_2^2$. 
In our implementation, we generate $M$ different Langevin samples $\{\bar{\bbx}_m\}_{m=1}^M$ for each pair $\{\bby, \bbH\}$ by running Algorithm~\ref{alg} multiple times and keep the sample that minimizes
\begin{equation}\label{eq:Ntraj}
    \hat{\bbx} = \argmin_{\bbx \in \{\bar{\bbx}_m\}_{m=1}^M} ||\bby - \bbH\bbx||_2^2.
\end{equation}
\vspace{-0.3cm}

\noindent Notice that these $M$ Langevin trajectories can be run in parallel, as they are independent of each other.

\vspace{1mm}
\noindent {\bf Computational complexity.} 
The first step in Algorithm~\ref{alg} is to compute the SVD of the channel $\bbH$, whose complexity is 
\linebreak$\ccalO(N_uN_r\min\{N_u, N_r\})$, and is done only once per channel.
Then, the discretization schemes entails three steps.
The first one as well as the third are just vector summations, since we consider the mass parameter $\bbM = m \bbI$ to be just a scalar in this work. 
The second steps required the computation of~\eqref{eq:full_score}, which entails a complexity of $\ccalO(N_u^2 + KN_u)$ per iteration.
Therefore, the overall complexity, including the SVD computation and all the iterations, is $\ccalO(N_uN_r\min\{N_u, N_r\} + LT(N_u^2 + KN_u))$.
Compared to the overdamped case (see~\cite{zilberstein2022annealed}), we see that we are not adding computational burden to the detector.
Regarding the $M$ trajectories, observe that these are independent of each other, so they can be computed in parallel.
In Section~\ref{S:numerical_experiments}, we present some numerical experiments that analyze this trade-off and the impact on the SER performance.

\vspace{-0.05in}
\section{Numerical Experiments}
\label{S:numerical_experiments}

\vspace{-0.05in}
In this section, we present the results of our proposed method.\footnote{Code to replicate the numerical experiments can be found at \url{https://github.com/nzilberstein/Langevin-MIMO-detector}}
We start by presenting the channel model and the simulation setup.
In the first experiment, we analyze the SER performance of the proposed method as a function of the signal-to-noise ratio (SNR) and compare it to the overdamped Langevin detector for different noise levels $L$.
In the second experiment, we compare our proposed method with both classical and learning-based baseline detectors.
Finally, we compare the SER of the underdamped Langevin-based detector with respect to the other baselines as a function of the running time for a particular SNR.

\vspace{1mm}
\noindent {\bf Channel model and simulation settings.} The channel model is generated following the Kronecker correlated model
\begin{equation}
	\bbH = \bbR_r^{1/2}\bbH_e \bbR_u^{1/2},
	\label{E:kron}
	\vspace{-0.07in}
\end{equation}
where $\bbH_e$ is a Rayleigh fading channel matrix and $\bbR_r$ and $\bbR_u$ are the spatial correlation matrices at the receiver and transmitters, respectively, generated according to the exponential correlation matrix model with a correlation coefficient $\rho$; see \cite{Loyka2001} for details.
The SNR is given by 
\begin{equation}
	\text{SNR} = \frac{\mathbb{E}[||\bbH\bbx||^2]}{\mathbb{E}[||\bbz||^2]} = \frac{N_u}{\sigma_0^2N_r}.
\end{equation}
The simulation environment includes a base station with $N_r=64$ receiver antennas and $N_u=32$ single-antenna users.
We consider a 16-QAM modulation and $\rho = 0.6$. 
The value of $\epsilon$ is fixed at $6 \times 10^{-4}$ and the number of samples per noise level at $T=30$ unless otherwise specified.
We set $\gamma = 1$ and the mass is defined as $\bbM = \frac{\gamma^2}{4} \bbI$.
The batch size for testing is $5000$. 

\vspace{1mm}
\noindent{\bf Comparison with the overdamped Langevin detector.}
In this experiment, we compare the advantage of using the underdamped Langevin detector w.r.t. the overdamped case. 
We consider three settings: the first one is with $L = 5$, $M = 20$, $\tau = 0.01$, $\sigma_1 = 0.4$, and $\sigma_5=0.02$, which corresponds to the \emph{first low computational complexity regime (low 1)}, while the \emph{second low complexity one (low 2)} is with $L = 5$, $\sigma_1 = 1$, $\sigma_5=0.01$, $\epsilon = 3\times 10^{-5}$, $M = 20$, $T = 30$ and $\tau = 0.1$.
The third setting is the \textit{high computational complexity (high)}, that corresponds to the same configuration as in~\cite{zilberstein2022annealed}, and is like low 2 but with $L = 20$, $T = 70$ and $\tau = 0.5$.
The results are shown in Fig.~\ref{fig:SER-comparison_order}.
First, we observe that in the high computational complexity regime both methods have the same performance. 
On the other hand, in both low regimes, the underdamped Langevin detector outperforms the overdamped Langevin detector, where the number of iterations ($L \times T$) was reduced by a factor of 10 (150 vs 1400).
Therefore, this illustrates the trade-off between performance and running time.

\vspace{1mm}
\noindent{\bf Performance comparison with baseline methods.}
Based on our previous experiments, we consider the \emph{low 1} regime for the underdamped and \emph{low 2} for the overdamped, and the following baseline detectors: MMSE detector, V-BLAST detector~\cite{V-blast}, overdamped Langevin with 5 levels, and two learning-based methods, RE-MIMO~\cite{remimo} and OAMPNet~\cite{oampnet}, which were trained as explained in the respective papers with channels drawn from~\eqref{E:kron}.
The comparison is shown in Fig.~\ref{fig:SER-comparison}.
The figure reveals that our proposed method markedly outperforms the other detectors (we omit the case of $L=20$ as we focus on low-complexity schemes).
Notice that our method can handle a varying number of users without the need for any retraining as required in, e.g., OAMPNet~\cite{oampnet}.
This is relevant in MIMO communications, as the number of users connected to the network is constantly changing.

\vspace{1mm}
\noindent{\bf Running time comparison.} 
In this third experiment, we compute the SER of our method and the other baselines w.r.t. running time in $\mathrm{ms}/\mathrm{symb}$.
We assume a coherence time such that each block contains 1000 samples.
The comparison is shown in Fig.~\ref{fig:SER-complexity} for an $\text{SNR}= \SI{16}{\decibel}$, $M=1$, and 5000 symbols.
Given the coherence time, we have to compute 5 SVDs that correspond to each $\{\bbH_i\}_{i=1}^5$.
First, notice that both underdamped and overdamped have the same performance when considering $L = 20$, something expected given the result in the first experiment.
However, when $L =5$, the underdamped case successfully reduces the running time while achieving a better performance compared to the overdamped case.

%
\vspace{-0.1in}
\section{Conclusions}
\vspace{-0.1in}
We proposed a general framework for linear inverse problems based on an annealed version and a discretization based on a splitting technique of the {underdamped} Langevin dynamics.
We applied it to the problem of MIMO detection and show that our proposed detector outperforms other methods including the overdamped Langevin detector in the low computational complexity regime.
Future work includes studying higher-order Langevin dynamics, like the third-order method proposed in~\cite{high_order_langevin}.


\bibliographystyle{IEEEbib}
\bibliography{citations}

\end{document}